\documentclass[12pt]{article}
\usepackage{graphicx}
\newcommand{\be}{\begin{equation}}
\newcommand{\ee}{\end{equation}}
\newcommand{\bea}{\begin{eqnarray}}
\newcommand{\eea}{\end{eqnarray}}
\newcommand{\al}{\\[0.4cm]}
\title{\Large \bf SUMMATION AND RENORMALIZATION OF BUBBLE GRAPHS
TO ALL ORDERS}
\author{\normalsize H. VERSCHELDE \\
\it Universiteit Gent \\
\it Vakgroep Wiskundige Natuurkunde en Sterrenkunde \\
\it Krijgslaan 281 - S9 \\
\it B-9000 Gent, Belgium}
\begin{document}
\maketitle
\vskip 36pt
\begin{abstract}

We introduce the 2PPI expansion which sums the bubble graphs
to all orders. We show that this expansion can be renormalised
with the usual counterterms in a mass independent scheme.
We discuss its application to the O(N) linear $\sigma$-model.
\end{abstract}

\newpage

Over the years, the summation of bubble graphs has attracted a lot of
interest because of the need to resort to non-perturbative approaches
in quantum field theory at finite temperature [1]. In het O(N) linear
$\sigma$-model, bubble graphs were originally [2] summed using the 
$N \rightarrow \infty$ approach. For finite N, it took some while to
overcome combinatorial problems [3], but finally consistent results were
obtained in [4,5], using the CJT [6] formalism. There were still some problems
though with renormalization which could only be implemented consistently
in the $N \rightarrow \infty$ limit.
\al
In this letter we discuss the 2PPI expansion which we introduced in [7],
as an approximation scheme for calculating the effective 
potential of the local composite operator $\phi^2$.
It resembles the 2PI expansion or CJT-method, in that it restricts the
topology of the Feynman diagrams at the cost of introducing a selfconsistency
condition. In contradistinction to the 2PI expansion, the self consistency
equation (gap equation) is local for the 2PPI expansion.
First, we will rederive
the 2PPI expansion in a new way which makes the renormalization proof more 
transparent. Then we will show that the usual counterterms renormalize the 2PPI
expansion when using a mass independent scheme. Next we calculate the one
loop renormalised 2PPI result for the effective potential in the O(N) linear
$\sigma$-model and show that it yields the renormalised sum of daisy and
superdaisy graphs. Finally we discuss the renormalization problems encountered
in previous treatments and show how they are neatly solved in the 2PPI approach.
\al
We will introduce the unrenormalized 2PPI expansion using the
$\lambda \phi^4$ theory for simplicity. Generalization to the
O(N) linear $\sigma$-model is straightforward. We start from the
1PI expansion and try to sum all seagull and bubble graphs
exactly. These insertions arise in 2PPR or two particle point 
reducible graphs
because they disconnect from the rest of the diagram when two
lines meeting at the same point (the 2PPR-point) are cut
(fig.1). We notice that the seagull and bubble graphs
contribute to the self energy as effective masses $\frac{\lambda}
{2}\varphi^2$ and $\frac{\lambda}{2} \langle \phi^2\rangle_c =
\frac{\lambda}{2} \Delta$ respectively. Therefore, we could try
to sum all 2PPR insertions by simply deleting the 2PPR graphs
from the 1PI expansion and introducing the effective mass :
\be
\overline{m}^2 = m^2 + \frac{\lambda}{2} (\varphi^2 + \Delta)
\ee
in the remaining 2PPI graphs. This is too naive though since
there is a double counting problem which can be easily
understood in the simple case of a 2loop vacuum diagram (daisy
graph with two petals) of fig. 2.a. Each petal can be seen as a
selfenergy insertion in the other, so there is no way of
distinguishing one or the other as the remaining 2PPI part. We
could however earmark one of the petals by applying a derivative
with respect to $\varphi$ (fig. 2.b). This way the 2PPI
remainder (which contains the earmark) is uniquely fixed. Now,
there are two ways in which the derivative can hit a $\varphi$
field. It can hit an explicit $\varphi$ field which is not a
wing of a seagull or it can hit a wing of a seagull or implicit
$\varphi$ field hidden in the effective mass. We therefore have
: 
\bea
\frac{\delta}{\delta \varphi} \Gamma_q^{1PI}(m^2,\varphi) 
& = & \frac{\partial}{\partial \varphi}
\Gamma_q^{2PPI}(\overline{m}^2,\varphi) \nonumber \\
& + & \lambda\varphi \frac{\partial}{\partial \overline{m}^2}
\Gamma^{2PPI}_q (\overline{m}^2,\varphi) 
\eea
where $\Gamma^{1PI} = S(\varphi) + \Gamma_q^{1PI}$ or using
(1) :
\be
\frac{\delta}{\delta\varphi} \Gamma_q^{1PI}(m^2,\varphi) =
\frac{\delta}{\delta \varphi}
\Gamma_q^{2PPI}(\overline{m}^2,\varphi) - \frac{\lambda}{2}
\frac{\delta \Delta}{\delta \varphi} \frac{\partial
\Gamma^{2PPI}_q}{\partial \overline{m}^2}
(\overline{m}^2,\varphi) 
\ee
Using an analogous combinatorial argument, we have :
\be
\frac{\partial \Gamma^{1PI}_q}{\partial m^2} (m^2,\varphi) =
\frac{\partial \Gamma^{2PPI}_q}{\partial \overline{m}^2}
(\overline{m}^2,\varphi) 
\ee
and since $\Delta/2 = \frac{\partial \Gamma_q^{1PI}}{\partial
m^2}$ we find the following gap equation for $\Delta$ :
\be
\frac{\Delta}{2} = \frac{\partial \Gamma^{2PPI}_q}{\partial
\overline{m}^2} (\overline{m}^2,\varphi)
\ee
The gap equation (5) can be used to integrate (3) and we finally
obtain :
\be
\Gamma^{1PI}(m^2,\varphi) = S(\varphi) + \Gamma^{2PPI}_q (m^2 +
\frac{\lambda}{2} (\varphi^2 + \Delta),\varphi) -
\frac{\lambda}{8} \int d^Dx \Delta^2
\ee
This equation together with the gapequation (5) sums the
seagulls and bubbles to all orders. The first few diagrams of
the 2PPI expansion are displayed in fig.3.
\al
To be useful, we have to show that equation (6) which relates
1PI and 2PPI expansions can be renormalised. Again, we earmark
the 1PI diagrams by applying a $\varphi$ derivative so that 2PPR
and 2PPI parts are unambiguous. We first renormalize the bubble
subgraphs. Consider a generic bubble inserted at the 2PPR point
x (fig. 4.a). All primitively divergent subgraphs of the bubble
graph which do not contain the 2PPR point x can be renormalised
with the counterterms of the $\lambda \phi^4$ Lagrangian :
\be
\delta {\cal L} = \frac{\delta Z}{2} (\partial_{\mu}
\phi)^2 + \delta Z_2 \frac{m^2\phi^2}{2} + \delta Z_{\lambda}
\frac{\lambda \phi^4}{4!}
\ee
As a consequence of these substractions, the contribution of the
bubbles to the effective mass is $\frac{\lambda}{2} \langle \phi^2\rangle_c$
where the connected V.E.V. is now calculated with the full
Lagrangian, counterterms included. For subgraphs of the bubble
containing the 2PPR point, we need only the 2PPR-parts of the
counterterms. Let's first renormalise the proper subgraphs of the
bubble which contain x. Their generic topology is displayed in
fig. 4.b and fig. 4.c. They can be made finite with the 2PPR
part $\delta Z_{\lambda}^{2PPR} \lambda \phi^4/4!$ and
contribute $\frac{\lambda}{2} \delta Z^{2PPR}_{\lambda}
\langle \phi^2\rangle_c$ to the effective mass. We still have to subtract the
overall divergences of the bubble graph. Their generic topology
is displayed in fig. 4.d and 4.e for coupling constant
renormalization and fig. 4.f for mass renormalization. Again
only the 2PPR parts of the counter\-terms have to be used and the
subtraction of the overall divergences contributes
$\frac{\lambda}{2} \delta Z^{2PPR}_{\lambda} \varphi^2 + \delta
Z_2^{2PPR} m^2$ to the effective mass. Because we use mass
independent renormalisation the 2PPR-part of coupling constant
renormalisation $\delta Z^{2PPR}_{\lambda}$ can be related to
mass renormalisation $\delta Z_2$.
Indeed, the divergent parts of the subgraphs in fig. 4.d and
4.e which are of the coupling constant type can be viewed as
divergent mass renormalizations with a mass insertion at the
2PPR point $x$ and the $\varphi$ legs as incoming and outgoing
legs of the selfenergy graph. Because we use a mass independent
renormalization scheme (for example minimal substration), not
only the divergent parts but also the finite parts of the
corresponding counterterms can be chosen equal so that we have
\be
\delta Z^{2PPR}_{\lambda} = \delta Z_2
\ee
Adding the various contributions coming
from renormalising the bubble graphs, we find for the
renormalised effective mass :
\be
\overline{m}^2_R = m^2 \left( 1 + \delta Z_2^{2PPR}\right) + \frac{\lambda}{2}
\left( 1 + \delta Z_{\lambda}^{2PPR}\right) (\varphi^2 + \Delta)
\ee
where $\Delta = \langle \phi^2 \rangle_c$ \footnote{The
composite operator V.E.V. $\Delta = \langle \phi^2 \rangle_c$ is
different here from the unrenormalised one in the first
paragraph where it is calculated with the Lagrangian without counterterms.} 
or using (8) :
\be
\overline{m}^2_R = m^2 + \frac{\lambda}{2} \left(Z_2(\varphi^2 +
\Delta) + \frac{2\delta Z^{2PPR}_2}{\lambda} m^2 \right)
\ee
We can rewrite this as :
\be
\overline{m}^2_R = m^2 + \frac{\lambda}{2} (\varphi^2 +
\Delta_R) 
\ee
where we have introduced the renormalised V.E.V. of the composite
operator $\phi^2$~:
\be
\langle \phi^2 \rangle_{c,R} = \Delta_R = Z_2 \Delta + \delta Z_2 \varphi^2 +
\frac{2\delta Z^{2PPR}_2}{\lambda} m^2
\ee
After renormalizing the bubble subgraphs, the $\varphi$
derivative of $\Gamma^{1PI}_q$ can be written as :
\be
\frac{\delta}{\delta \varphi} \Gamma^{1PI}_{q,BR}(m^2,\varphi) =
\frac{\partial}{\partial \varphi} \Gamma^{2PPI}_q
(\overline{m}^2_R,\varphi) + \lambda\varphi
\frac{\partial}{\partial \overline{m}^2_R}
\Gamma_q^{2PPI}(\overline{m}^2_R, \varphi)
\ee
where BR stands for "bubble renormalised".
\al
Because there is no overlap, having renormalised the bubbles, we
can now renormalize the 2PPI remainder (which contains the
earmarked vertex). Let us first consider mass renormalization. A
subgraph $\gamma$ in the 2PPI remainder of
$\frac{\delta}{\delta\varphi} \Gamma^{1PI}_{q,BR}$ that needs
mass renormalization can be made finite with a counterterm
$\delta Z_2(\gamma)m^2 \phi^2/2$. However, for any such subgraph
$\gamma$, there are subgraphs $\gamma^{\prime}$ obtained from $\gamma$ by
replacing the mass $m^2$ with a seagull or renormalised bubble.
These subgraphs require coupling constant renormalization which
entails an effective counterterm $\delta
Z^{2PPR}_{\lambda}(\gamma^{\prime})(\frac{\lambda}{2}\varphi^2 +
\frac{\lambda}{2} \Delta_R)\phi^2/2$.
Taking into account the identity of the renormalization
constants for mass and 2PPR-coupling constant renormalization
(eq.8), the counterms for these mass-type divergent subgraphs of
$ \frac{\delta \Gamma^{1PI}_{q,BR}}{\delta \varphi}$ add up to
$\delta Z_2(\gamma)\overline{m}^2_R \phi^2/2$, exactly what is
needed for mass renormalization of
$\Gamma^{2PPI}_q(\overline{m}^2_R,\varphi)$ in the right hand
side of equation (13). The remaining divergent subgraphs need
wavefunction renormalization or are of the coupling constant
renormalization type that cannot be generated by inserting
seagulls or bubbles in a mass-type divergent subgraph. They are
made finite by counterterms independent of mass and hence are
the same for left and right hand sides of equation (13).
Therefore, we can conclude that in a mass independent
renormalization scheme, equation (2) can be renormalised with
the available counterterms as :
\be
\frac{\delta}{\delta \varphi} \Gamma^{1PI}_{q,R}(m^2,\varphi) =
\frac{\partial}{\partial \varphi}
\Gamma^{2PPI}_{q,R}(\overline{m}^2_R,\varphi) + \lambda \varphi
\frac{\partial}{\partial \overline{m}^2_R}
\Gamma^{2PPI}_{q,R}(\overline{m}^2_R, \varphi)
\ee

To proceed, we have to renormalise the gap equation (eq.4).
Using essentially the same arguments as in the previous
paragraphs, we find that
\be
\frac{\partial \Gamma^{1PI}_{q,R}}{\partial m^2}(m^2,\varphi) =
\frac{\partial \Gamma^{2PPI}_{q,R}}{\partial \overline{m}^2_R}
(\overline{m}^2_R, \varphi)
\ee
From the pathintegral we readily obtain
\be
\frac{\partial \Gamma^{1PI}_R}{\partial m^2} (m^2,\varphi) =
\frac{1}{2} Z_2 (\varphi^2 + \langle \phi^2 \rangle_c) +
\frac{\partial}{\partial m^2} (\delta E_{vac})
\ee
where $ \delta E_{vac}$ is the counterterm needed to cancel
divergences in the vacuum energy. In dimensional regularisation,
one can easily show that $\delta E_{vac} = \frac{m^4}{2} \delta
\zeta$, where $\delta \zeta$ is the logarithmically divergent
part of $\langle \phi^2 (x)\phi^2(y)\rangle$ which, up to finite parts, is
equal to $\frac{\delta Z^{2PPR}_2}{\lambda}$. In a mass
independent renormalization scheme, finite renormalization can
be chosen so that
\be
\delta E_{vac} = \frac{m^4}{2\lambda} \delta Z^{2PPR}_2
\ee
Using $\frac{\partial}{\partial m^2} \Gamma^{1PI}_R = \varphi^2/2 + \frac
{\partial}{\partial m^2} \Gamma^{1PI}_{q,R}$
and equations (15), (16) and (17) we obtain
\be
\frac{1}{2} \left( Z_2 \langle \phi^2 \rangle_c + \delta Z_2 \varphi^2 + 2m^2
\frac{\delta Z_2^{2PPR}}{\lambda}\right) = \frac{\partial
\Gamma^{2PPI}_{q,R}}{\partial \overline{m}^2_R}
(\overline{m}^2_R,\varphi) 
\ee
which because of equation (12) can be written as the
renormalised gap equation~:
\be
\frac{\Delta_R}{2} = \frac{\partial
\Gamma^{2PPI}_{q,R}}{\partial \overline{m}^2_R}
(\overline{m}^2_R,\varphi) 
\ee
As in the unrenormalised case (cfr. eq.3), this gapequation can
be used to integrate equation (14) and we finally arrive at :
\be
\Gamma^{1PI}_R (m^2,\varphi) = S(\varphi) + \Gamma^{2PPI}_{q,R}
\left( m^2 + \frac{\lambda}{2} (\varphi^2 + \Delta_R),\varphi \right) -
\frac{\lambda}{8} \int d^D x \Delta^2_R
\ee
This equation together with the gapequation (19) sums the
seagulls and renormalised bubbles to all orders.
To renormalize $\Gamma^{1PI}$, it is therefore sufficient to renormalize $\Gamma^{2PPI}$
using a mass independent renormalisation scheme (MS for example),
calculate the renormalised bubble $\Delta_R$ from the gapequation
and substitute in eq.(20).
\al
The preceding results can be easily generalised to the O(N) linear $\sigma$-model with 
Lagrangian :
\be
{\cal L} = \frac{1}{2} \partial_{\mu} \phi_i \partial_{\mu}\phi_i +
\frac{m^2_{ij}}{2} \phi_i \phi_j + \frac{\lambda}{8} (\phi_{ii})^2
\ee
Seagulls and bubbles can be included in an effective mass
\be
\overline{m}^2_{ij} = m^2_{ij} + \lambda [\varphi_i \varphi_j + \Delta_{ij}] + 
\frac{\lambda}{2} [\varphi^2 + \Delta _{kk}]\delta_{ij}
\ee
where $\Delta_{ij} = \langle \phi_i \phi_j\rangle_c$. For the O(N)-linear $\sigma$-model,
equations (2) and (3) become~:
\bea
\frac{\delta}{\delta \varphi_k} \Gamma^{1PI}_q (m^2,\varphi) 
& = & \frac{\partial}{\partial
\varphi_k} \Gamma^{2PPI}_q (\overline{m}^2,\varphi) + [\lambda \varphi_k \delta_{ij} +
\lambda(\delta_{ik} \varphi_j + \delta_{jk} \varphi_i)] \frac{\partial \Gamma^{2PPI}_q}
{\partial \overline{m}^2_{ij}} (\overline{m}^2,\varphi) \nonumber \\
& & \nonumber \\
& = & \frac{\delta}{\delta \varphi_k} \Gamma^{2PPI}_q (\overline{m}^2,\varphi) - \left[ \lambda
\frac{\delta \Delta_{ij}}{\delta \varphi_k} + \frac{\lambda}{2} \delta_{ij} \frac
{\delta \Delta_{ll}}{\delta \varphi_k}\right] \frac{\partial \Gamma^{2PPI}_q}
{\partial \overline{m}^2_{ij}} (\overline{m}^2,\varphi)
\eea
We can use the gapequation $\Delta_{ij}/2 = \frac{\partial \Gamma^{2PPI}}{\partial
\overline{m}^2_{ij}} (\overline{m}^2,\varphi)$ to integrate the last equation and obtain :
\be
\Gamma^{1PI}(m^2,\varphi) = S(\varphi) + \Gamma^{2PPI}_q (\overline{m}^2,\varphi) - 
\frac{\lambda}{8} \int d^D x \left[ (\Delta_{ii})^2 + 2(\Delta_{ij})^2 \right]
\ee
For $m^2_{ij} = \delta_{ij} m^2$, we can make use of O(N) symmetry to define :
\be
\overline{m}^2_{ij} = \frac{\varphi_i \varphi_j}{\varphi^2} \overline{m}^2_{\sigma} +
\left( \delta_{ij} - \frac{\varphi_i \varphi_j}{\varphi^2}\right) \overline{m}^2_{\pi}
\ee
and
\be
\Delta_{ij} = \frac{\varphi_i \varphi_j}{\varphi^2} \Delta_{\sigma} + \left(
\delta_{ij} - \frac{\varphi_i \varphi_j}{\varphi^2}\right) \Delta_{\pi}
\ee
so that because of (22) :
\bea
\overline{m}^2_{\sigma} & = & m^2 + \frac{3\lambda}{2}\left[ \varphi^2 + \Delta_{\sigma}
+ \frac{N-1}{3} \Delta_{\pi}\right] \nonumber \\
\overline{m}^2_{\pi} & = & m^2 + \frac{\lambda}{2} \left[ \varphi^2 + \Delta_{\sigma} + (N+1)
\Delta_{\pi}\right]
\eea
The relation between 1PI and 2PPI expansion now simplifies to
\bea
\Gamma^{1PI}(m^2,\varphi) & = & S(\varphi) + \Gamma^{2PPI}_q (\overline{m}^2_{\sigma},
\overline{m}^2_{\pi},\varphi) \nonumber \\
& - & \frac{\lambda}{8} \int d^D x \left[ 3 \Delta^2_{\sigma} + (N^2 - 1) \Delta^2_{\pi} +
2(N-1) \Delta_{\sigma} \Delta_{\pi}\right]
\eea
and the gap equations are
\bea
\frac{\delta \Gamma^{2PPI}}{\delta \overline{m}^2_{\sigma}} & = & \frac{\Delta_{\sigma}}{2}
 \nonumber \\
\frac{\delta \Gamma^{2PPI}}{\delta \overline{m}^2_{\pi}} & = & (N-1) \frac{\Delta_{\pi}}{2}
\eea

To show the efficiency of the 2PPI expansion, let's calculate the effective potential at one 
loop
2PPI. We choose $\varphi_i = \delta_{iN} \varphi$. Since there are $(N-1)$ masses 
$\overline{m}_{\pi}$, and one mass $\overline{m}_{\sigma}$ running in the loop, we have :
\be
V^{2PPI}_q (\overline{m}^2_{\sigma},\overline{m}^2_{\pi},\varphi) = \frac{1}{2} \int
\frac{d^D p}{(2\pi)^D} \ln (p^2 + \overline{m}^2_{\sigma}) + \frac{(N-1)}{2}
\int \frac{d^D p}{(2\pi)^D}\ln (p^2 + \overline{m}^2_{\pi})
\ee
The renormalization procedure of $\Gamma^{1PI}$ is completely analogous as in the $N=1$ case 
(for
details, see [8]). We can simply renormalize $\Gamma^{2PPI}$ using for example the
$\overline{MS}$ scheme and calculate the renormalised bubbles $\Delta_{\sigma,R}$ and
$\Delta_{\pi,R}$ by using the renormalised gapequations (29). We finally obtain
at one loop 2PPI :
\bea
V_R^{1PI}(m^2,\varphi) & = & V(\varphi) + \frac{\overline{m}^4_{\sigma,R}}{64\pi^2}
\left[ \ln \frac{\overline{m}^2_{\sigma,R}}{\overline{\mu}^2} - \frac{3}{2}\right]
+ (N-1) \frac{\overline{m}^4_{\pi,R}}{64\pi^2} \left[ \ln \frac{\overline{m}^2_{\pi,R}}
{\overline{\mu}^2} - \frac{3}{2}\right] \nonumber \\
& - & \frac{\lambda}{8} \left[ 3 \Delta^2_{\sigma,R} + (N^2 -1)\Delta^2_{\pi,R} + 2(N-1)
\Delta_{\sigma,R} \Delta_{\pi,R}\right]
\eea
where
\bea
\Delta_{\ast,R} = \frac{\overline{m}^2_{\ast,R}}{16\pi^2} \left[
\ln \frac{\overline{m}^2_{\ast,R}}{\overline{\mu}^2} -1 \right]
\eea
and $\overline{m}_{\sigma,R}$ and $\overline{m}_{\pi,R}$ are given by equation (27)
with renormalised $\Delta^{\prime}s$.
\al
From the topology of the one loop 2PPI graph, it is clear that our 1loop 2PPI result agrees 
with the Hartree approximation and should give the sum of daisy and superdaisy
graphs. If we compare our unrenormalised expression we find complete agreement
(after some algebra) with the unrenormalised sum of daisy and superdaisy
graphs published in [4,5] and [9] and obtained within the CJT formalism [6]
(2PI expansion). The advantage of our 2PPI expansion is that we quite naturally
obtain a simple expression by keeping only the one loop 2PPI vacuum bubble,
while in the 2PI approach one has to keep part of the 2loop graphs (the 
O($\lambda$) part) and the simple expression (31) is only obtained after some
rearrangement. Furthermore, one can easily calculate higher order terms in
the 2PPI expansion while in the 2PI expansion, it is very difficult to go
beyond the Hartree approximation because of the non-locality of the gap
equations. Concerning renormalization, we find complete agreement with the 
renormalised expression of [9] obtained using the auxiliary field method. We
disagree with the "non-perturbative" renormalisation used in [4,5]. These authors
find that in order to obtain a finite effective mass (which sums daisy and
superdaisy graphs), the coupling constant has to run in a different way than
is dictated by the perturbative renormalisation group. For the N=1 case, their
non-perturbative $\beta$-function is a factor of three smaller than the perturbative
one and is essentially the $N \rightarrow \infty$ $\beta$ function extrapolated
to N=1. Our 2PPI analysis to all orders can easily explain this paradox. We
showed in (9) that after adding counterterms, the renormalised effective mass
can be written as :
\be
\overline{m}^2_R = \left( m^{2PPR}_0 \right)^2 + \frac{\lambda^{2PPR}_0}{2}
(\varphi^2 + \Delta)
\ee
This relation is valid to all orders. At one loop one has 
$\delta Z^{2PPR}_2 = \delta Z_2 , \delta Z_{\lambda}^{2PPR} = \frac{1}{3}
\delta Z_{\lambda}$.
So, if one naively assumes $\lambda_0 = \lambda_0^{2PPR}$, one obtains a
$\beta$-function which is a factor of three too small. This factor of three
is due to crossing which changes a 2PPR coupling constant renormalisation
insertion into a 2PPI one. This also explains why the $N = \infty$ renormalisation
is straightforward : crossing terms (and hence Fock or 2PPI terms) are
subdominant and only the 2PPR or Hartree terms survive in the $N \rightarrow \infty$
limit.
\al
In summary, we have presented a method for summing and renormalising bubble
graph insertions to all orders based on the 2PPI expansion. Besides the O(N)
linear $\sigma$-model, also Q.E.D. and models with four-fermion couplings
or Yukawa couplings can be treated in this expansion. Finite temperature
effects can be easily included. In [9], we sum daisy and superdaisy graphs of
O(N) linear $\sigma$-model at finite T, using the 2PPI expansion. Renormalisation
is straightforward and the Goldstone theorem is obeyed to all orders. Extension
to the non-equilibrium domain is possible [10].

\begin{figure}
\centering
\includegraphics[width=10cm,height=10cm]{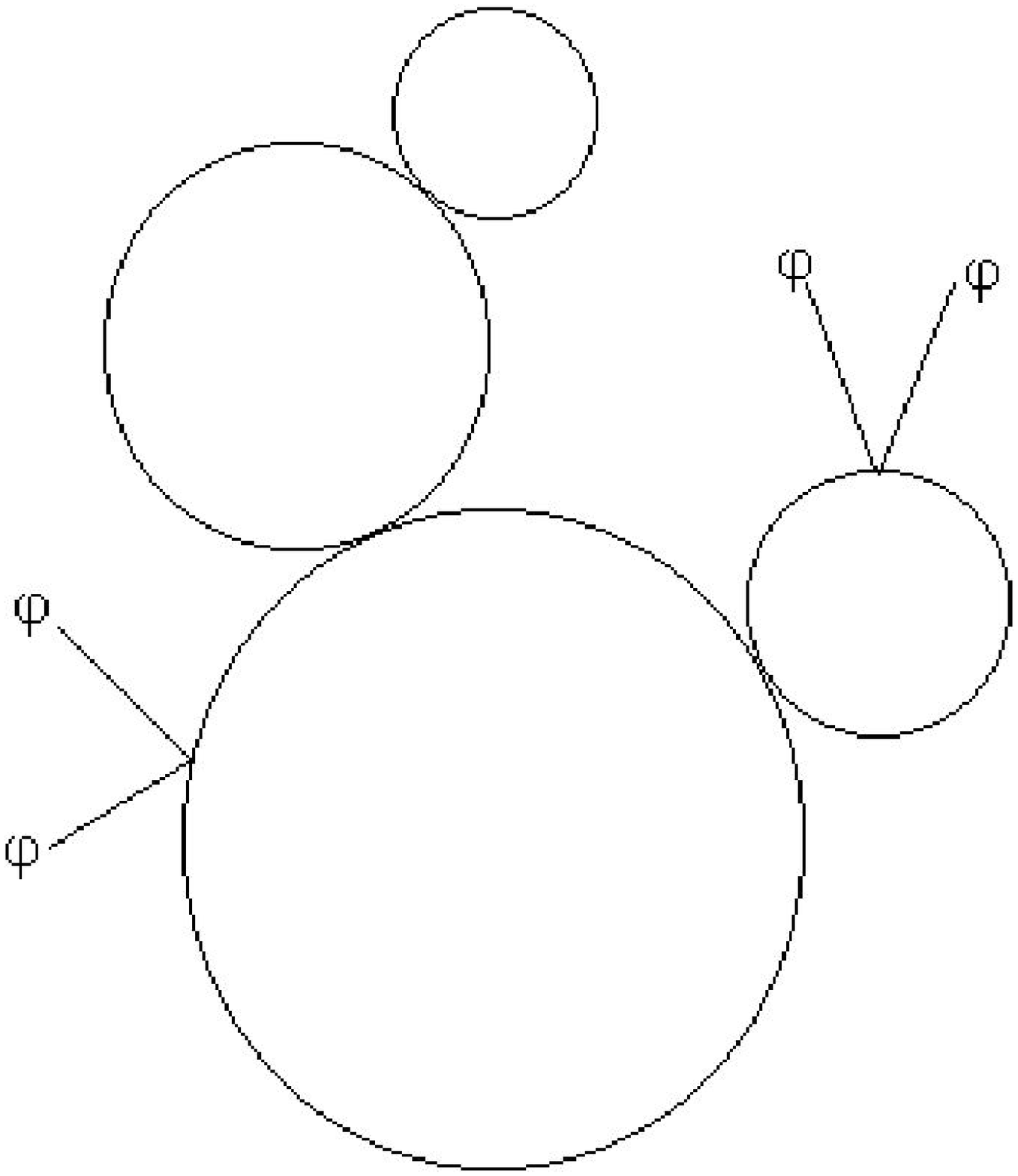}
\caption{Generic 2PPR diagram}
\end{figure}
\newpage
\begin{figure}
\centering
\includegraphics[width=10cm,height=12cm]{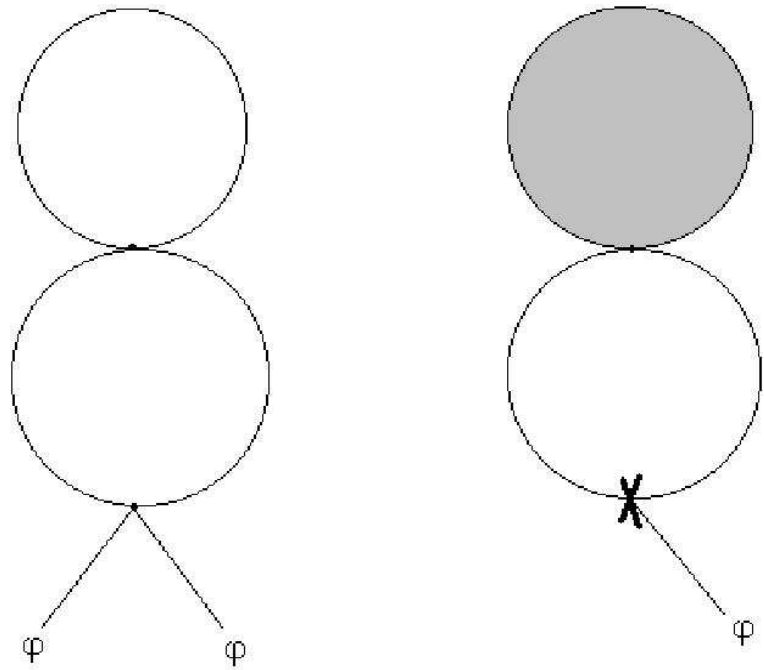}
\caption{2PPR part is shaded, 2PPI rest is earmarked}
\end{figure}
\newpage
\begin{figure}
\centering
\includegraphics[width=15cm,height=5cm]{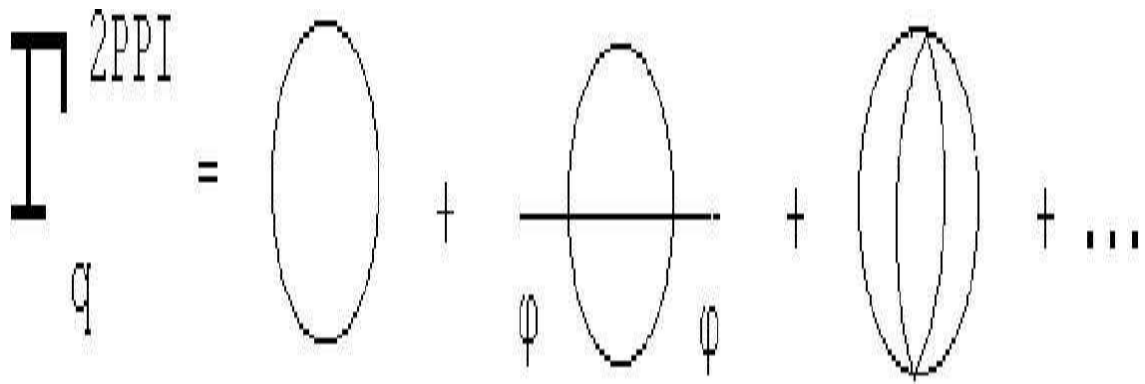}
\caption{First terms of 2PPI expansion}
\end{figure}
\newpage
\begin{figure}
\centering
\includegraphics[width=14cm,height=20cm]{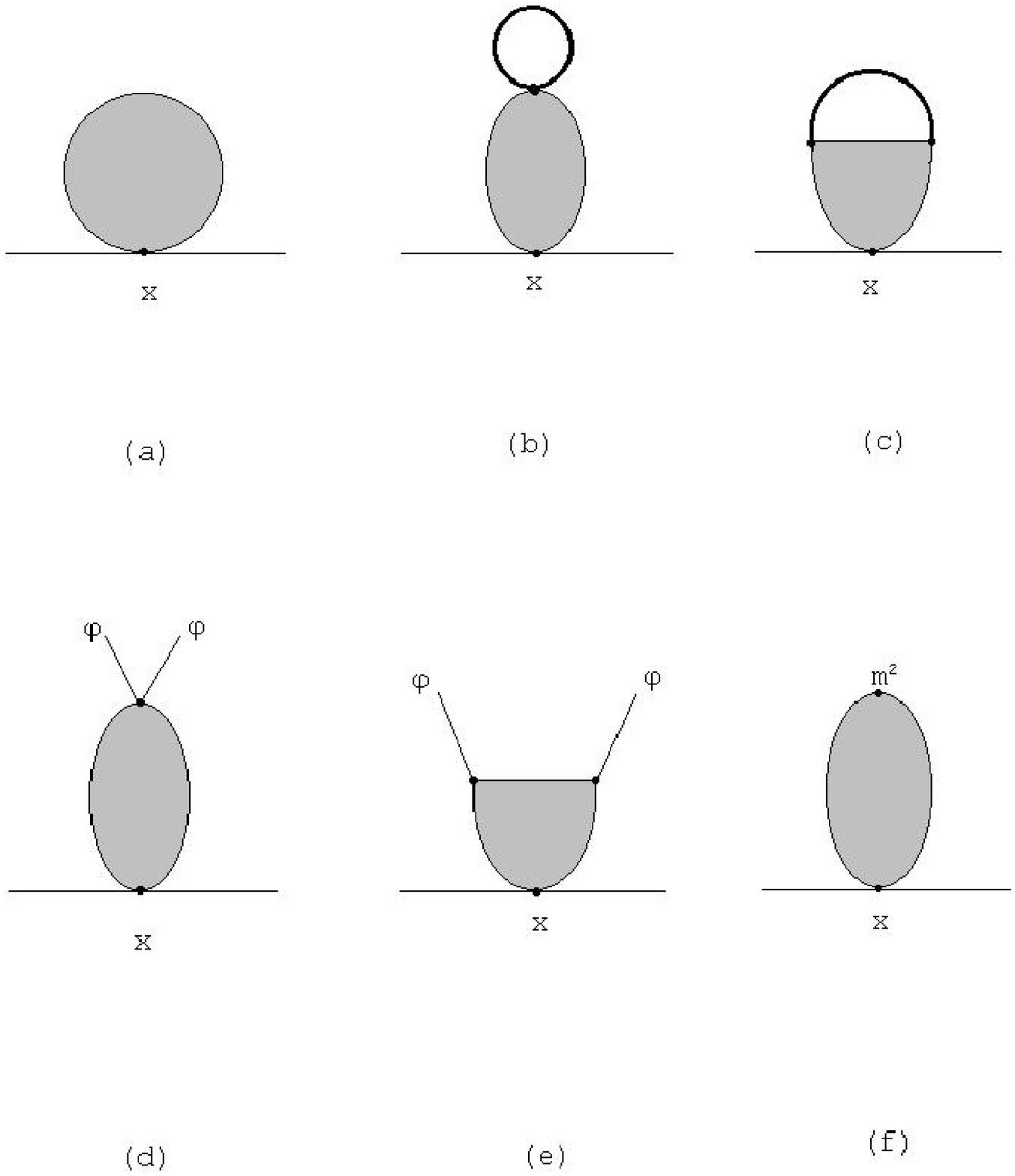}
\caption{Generic bubble(shaded) and its subdivergences(shaded).Thick lines are full 
propagators.}
\end{figure}
\end{document}